# Circulator based on spoof surface plasmon polaritons

Tianshuo Qiu, Yongfeng Li, Jiafu Wang, and Shaobo Qu

*Abstract*—Circulators based on spoof surface plasmon polaritons are designed and analyzed. In the letter, we use blade structure to realize the propagation of SSPPs wave and a matching transition is used to feed energy from coplanar waveguide to the SSPPs. And the circulator shows good nonreciprocal transmission characteristics. The simulation results indicate that in the frequency band from 5 to 6.6 GHz, the isolation degree and return loss basically reaches 15dB and the insertion loss is less than 0.5dB. Moreover, the use of confinement electromagnetic waves can decrease the size of the ferrite and show a broadband characteristic.

## I. Introduction

Three-port circulators are passive nonreciprocal devices, in which a microwave or radio frequency signal entering any port is transmitted only to the next port in rotation[1-5]. Nowadays, circulators are used more often in microwave area. For instance, circulators are frequently used in telecommunication networks, such as RADARs, cellular communications and broadcasting systems, to connect transmitters and receivers to the same antenna, avoiding interference between them [6]. In circulator, ferrites are usually used to realize the nonreciprocal characteristics. However, to get a broadband, junction circulator should have a long transmission line for impedance matching. And the junction circulator has a bulky ferrite. So making ferrite thin and small is still a challenge.

Recently, more efforts have been made towards the development of spoof surface plasmon polaritons (SSPPs) for microwave propagation. SSPPs is the high-confinement SPPs (surface plasmon polaritons) at low frequency, in which EM(electromagnetic) waves are bound on the metal/dielectric interface, propagating parallel to the interface and decaying exponentially in the direction vertical to the interface, just like "real SPPs" [7-10]. J. Pendry et al. reported that structured metal surface can support and propagate SSPPs [7]. In order to produce high transmission efficiency between single wire and coplanar waveguide (CPW), many efforts have been made [11–14]. As we know, SSPPs have good characteristics of high field confinement, planar configuration and deep-subwavelength, so we can take the advantage of the SSPPs to fabrication circulator.

In this paper, we study a microwave circulator based on SSPPs. We propose to utilize metallic blade structure to realize SSPPs and replace the stripline and junction with SSPPs. The addition of high-confinement electric field reduces the space to realize circulation performance and makes broad bandwidth. And a matching transition has been proposed, which is constructed by gradient corrugations and flaring ground, to match both the momentum and impedance of CPW and the SSPPs.

## II. Theory, and Design

The circulator is composed of an inner conductor which has three access. Above and below this inner conductor, there are two circular discs of ferrite. Dielectric sleeve (NETEC NY9208 $\varepsilon_r$=2.08 tan $\delta$ =0.001) is around the ferrite for matching then two ground planes are closing the structure, as presented in Fig. 1.

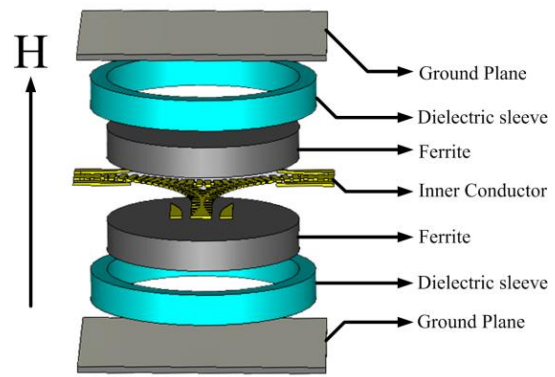

Fig. 1. Geometric model of the proposed SSPPs circulator

The material of the ferrite is yttrium iron garnet (YIG). Saturation magnetization (4πMs) is 1800Gauss, resonance linewidth is 15Oe and permittivity is 14.3. Ferrite with radius $r_0$=4.7 mm and height $h_0$=1.4mm is considered by theory analysis. In the letter, to realize the propagation of SSPPs wave, blade structure is used in the inner conductor, as shown in Fig. 2(a). In order to feed energies into the SSPPs, we use CPW and propose a smooth conversion between SSPPs and the CPW, as

Project is supported by the National Natural Science Foundation of China (Grants Nos. 61331005, 11204378, 11274389, 11304393, 61302023), the National Science Foundation for Post-doctoral Scientists of China (Grant Nos. 2013M532131, 2013M532221).

Tianshuo Qiu is with College of Science, Air Force Engineering University, Xi'an, Shaanxi 710051,China (e-mail: 909154790@qq.com).
Yongfeng Li is with College of Science, Air Force Engineering University, Xi'an, Shaanxi 710051,China (e-mail: liyf217130@126.com).
Jiafu Wang is with the College of Science, Air Force Engineering University, Xi'an, Shaanxi 710051,China (e-mail: wangjiafu1981@126.com).
Shaobo Qu is with College of Science, Air Force Engineering University, Xi'an, Shaanxi 710051,China (e-mail: qushaobo@mail.xjtu.edu.cn).

presented in Fig. 2(b). In the conversion section, the matching transition with gradient grooves and flaring ground are used to make conversion high-efficiency. The groove depth varies from $h_1$ =0.9mm to $h_3$= 2.7mm with a step of 0.9mm. The length and width of the flaring ground are designed as $L_2$ = 2.4 mm, $L_3$= 0.9 mm, respectively. The parameters for the blade structure are optimized to be: the period of the blade p=0.5mm, the width of the grooves a=0.2mm, and the depth of grooves h=0.2mm, the outer radius $R_2$ is 6.35mm, as presented in Fig. 2(c).

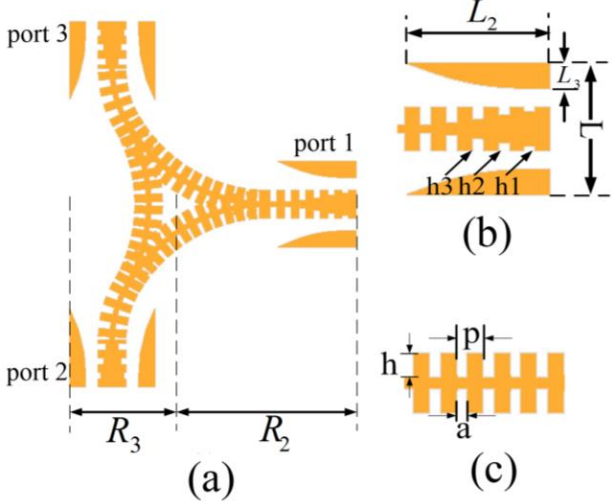

Fig. 2. Geometric model of the proposed inner conductor.
(a) Top view of inner conductor. $R_2$=6.35mm, $R_3$=3.65mm.
(b) The CPW section and the matching transition with gradient grooves and flaring ground, in which L=3mm, $L_2$ =2.4mm, $L_3$ =0.9mm, $h_1$ =0.09mm, $h_2$=0.18mm and $h_3$=0.27mm.
(c) Geometric of the metallic blade structure in detail, in which p= 0.5 mm, a=0.2 mm, and h = 0.4 mm.

For interconnect CPW to the SSPPs the smooth conversion section is employed. When the groove depth gradually increases and the ground of the matching transition is gradually flared out, the propagating wave number is changed from $k_0$ to $k_{spp}$, realizing the perfect momentum matching from CPW ($k_0$) to the blade structure ($k_{spp}$). As a result, the transmission efficiency of the metallic blade structure has enhanced and the impedance and momentum between SSPPs mode and CPW match perfectly.

As we all know, ferrite in a magnetic field shows anisotropy characteristic. To understand the origin of the anisotropy, the interaction between magnetic moments of the ferrite and a magnetic field needs to be studied. The spectral expression of the equation of motion leads to a complete permeability tensor

$$\mu = \begin{bmatrix} \mu & j\kappa & 0 \\ -j\kappa & \mu & 0 \\ 0 & 0 & \mu_0 \end{bmatrix} \quad (1)$$

Where $\kappa$ and $\mu$ are the tensor elements of the ferrite. The elements of the permeability tensor depend on the frequency and also depend on the magnitude of the applied field as well as on the saturation magnetization of the material.

When the high-confinement EM wave propagates along the blade structure, the constituents of both electric field and magnetic field are in many directions. If the medium around the blade structure is anisotropy such as ferrite, some constituents will interact with the medium and then the intensity will be changed. That will make intensity of electromagnetic wave in some section increase or decrease. In this case, we assume the energy is put in port 1. When the metallic blade structure is separated into different direction. Due to the anisotropy of the ferrite, the energy propagates from port 1 to port 2 instead of average distribution in the cross point. Obviously, the component will realize the circulation performance.

High-efficiency transmission can be realized as long as the propagating wave number has changed from $k_0$ to $k_{spp}$ by conversion section that interconnect CPW to the SSPPs. And the conversion section can realize wideband transmission by itself. So the quarter-wave transmission line for impedance matching is unnecessary to realize wideband performance. That makes the circulator smaller undoubtedly, because the conversion section that interconnect CPW to the SSPPs is much smaller than quarter-wave transmission line section. Moreover, less space are needed because of the characteristic of the high-confinement electromagnetic wave. The SSPPs wave are confined close to the blade structure. That can reduce the bulk of ferrite.

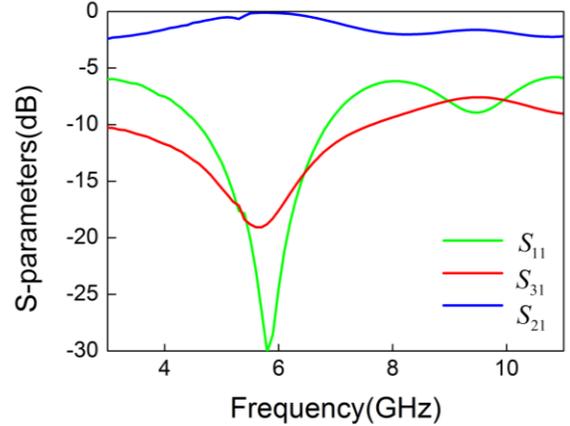

Fig. 3. The S parameters of the Circulator based on SSPPs.

### III. RESULT AND DISCUSSION

A high frequency electro-magnetic field simulation software based on a three dimensional finite element method, Ansoft HFSS 15.0, is used to analyze the transmission characteristics of the circulator. The external magnetic field we choose is 10000A/M. First of all, the simulated return loss, insertion loss, isolation, are studied and the results are shown in Fig. 3. From 5 GHz to 6.6 GHz, return loss and isolation degree basically reaches 15 dB, the insertion loss is less than 0.5dB. If we changed the design of the circulator, the bandwidth in which insertion loss is less than 1dB can reach 6 GHz (3GHz- 9GHz).

To verify the transmission efficiency of the conversion section, we simulate the conversion structure and blade structure, as shown in Fig.4. Figure 4 illustrates that the conventional guided waves in CPW section are turned to confinement electromagnetic wave successfully. And the SSPPs wave can propagate along the blade structure.

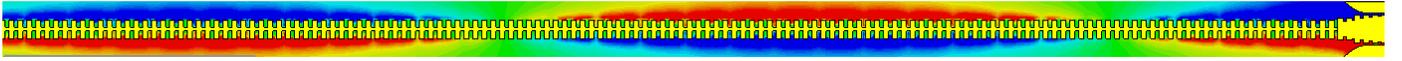

Fig.4. Simulated results of near electric field distributions in blade structure section and CPW section.

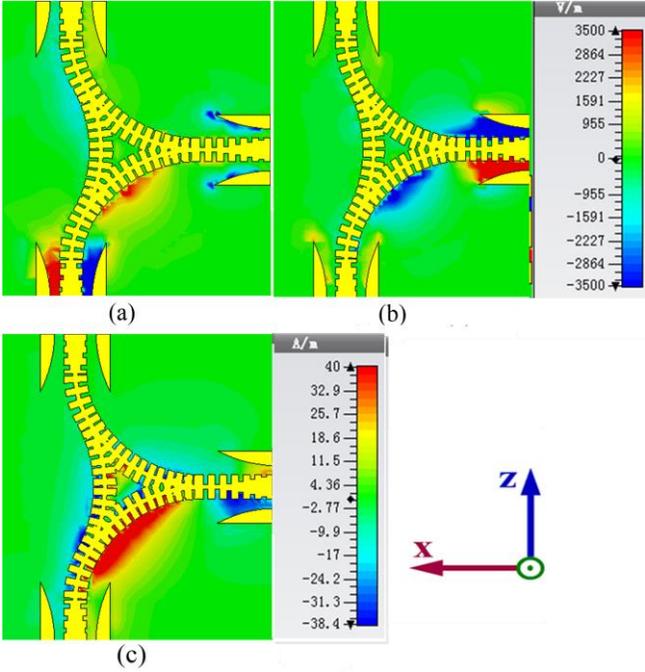

Fig.5. f=6GHz，the distribution of (a) the electric field x-component (b) the electric field y-component (c) the magnetic field y-component

Figure 5 shows the distribution of the electric field and magnetic field. We can clearly find that as the EM wave propagates along the conductor, the electric field z-component is transformed into x-component. And figure 6 shows the propagation of electric field in metallic blade structure of the circulator when energy is input in port 1. As predicted, a high-confinement electric field exists in SSPPs sections. The intensity of the confinement electric field are nearly the same in section Ⅰ and Ⅱ. As the confinement EM propagates along the X direction, the confinement electric field in section Ⅰ increases gradually. In the contrary, the confinement electric field in section Ⅱ decreases. That makes the circulation performance. We clearly observe the high-efficiency transmissions with very small reflections in the figure.

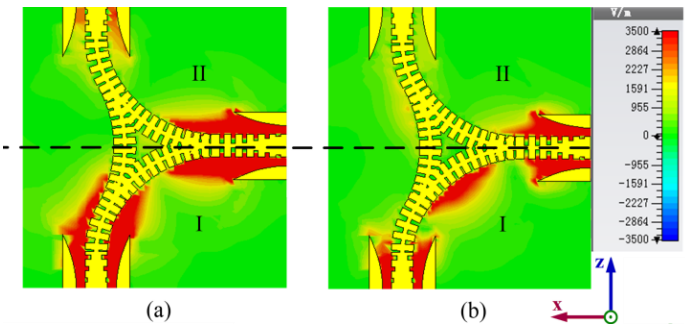

Fig.6. Simulated intensity distributions of the tangential electric field of the circulator (f=6GHz). (a) phase=50° (b) phase=110°

## IV. CONCLUSION

Through this paper we have applied the metallic blade structure to a [5-6.6] GHz band circulator, and design a high-performance circulator. We analyze the characteristics of the circulator. The results in the microwave frequency show the good characteristics of the circulator .This study proves that the circulator with SSPPs structure we designed operates with an acceptable nonreciprocal transmission characteristics as a practical circulator device.

In this letter, the confinement electromagnetic wave is lead into the new kind of circulator based on SSPPs. The confinement electromagnetic wave increases the characteristic of the circulator. The new kind of circulator doesn't need a long transmission line for impedance matching. And the structure can reduce the bulk of ferrite and then makes a smaller circulator. There are still lots of work to do to improve the characteristics. For example, we can utilize the center space without inner conductor to improve the bandwidth. And we can make the guided waves combine the SSPPs wave and electric and magnetic (TEM) mode to realize broadband characteristic.